\documentclass[twocolumn]{aa}  

\usepackage{graphicx}  

\def\B{{\em BeppoSAX }}  
  
\begin{document}  
  
\title{High-Energy pulse profile of the Transient X-ray  
  Pulsar SAX J2103.5+4545}    
\author{M. Falanga\inst{1}\fnmsep\thanks{\email{mfalanga@cea.fr}},  
  T. Di Salvo\inst{2} ,  
  L. Burderi\inst{3}, J. M. Bonnet-Bidaud\inst{1},  
  P. Goldoni\inst{1}, A. Goldwurm\inst{1}, G. Lavagetto\inst{2},  
  R. Iaria\inst{2}  and N. R. Robba\inst{2}}  
  
\offprints{M. Falanga}   
\titlerunning{High-energy properties of SAX J2103.5+4545}  
\authorrunning{M. Falanga, T. Di Salvo, et al.}  
  
\institute{CEA Saclay, DSM/DAPNIA/Service d'Astrophysique (CNRS FRE  
  2591), F-91191, Gif sur Yvette, France  
\and Dipartimento di Scienze Fisiche ed Astronomiche, Universit\`a  
   di Palermo, via Archirafi 36, 90123 Palermo, Italy  
\and Osservatorio Astronomico di Roma, via Frascati 33, 00040  
  Monteporzio Catone, Italy}

\abstract{  
  
In two recent {\em INTEGRAL} papers, Lutovinov et al. (2003) and  Blay et  
al. (2004) report a timing and spectral analysis of the transient  
Be/X-ray pulsar SAX J2103.5+4545 at high energies (5--200 keV).  
In this work we present for the first time a study of the pulse 
profile at energies above 20 keV  using {\em INTEGRAL} 
  data. The spin-pulse profile shows a prominent (with a duty  
cycle of 14\%) and broad (with a FWHM of $\sim$ 51 s) 
peak and a secondary peak which becomes more evident above 20 keV. The
pulsed fraction increases with energy from $\sim$ 45\% at 5--40 keV to
$\sim$ 80\% at 40--80 keV.   
The morphology of the pulse profile  also changes as a function of  
energy, consistent with variations in the spectral components that are
visible in the pulse phase resolved spectra. A study of the double 
peaked profile   
shows that the difference in the two peaks can be modeled by a  
different scattering fraction between the radiation from the two 
magnetic poles.   
  
 }   
\maketitle  
  
\section{Introduction}  
  
After the discovery by Hulleman et al. (1998) and 
suggestion that it is a Be X-ray binary, SAX J2103.5+4545 was 
conclusive confirmed as such by Reig et al. (2004) who extimate a 
distance of $\sim$ 6.5 kpc.    
The system consists of an accreting neutron star (NS) and a BOV  
main-sequence donor star which shows Balmer emission lines (Reig et al. 2004).  
The relatively short orbital period, 12.7 days (Baykal et  
al. 2000), and the  spin period of $\sim 358$ s (Hulleman et  
al. 1998)  make SAX J2103.5+4545 a peculiar Be/X-ray transient system.  
The source  does not follow the $P_{\rm pulse}$  
versus $P_{\rm orb}$ relation followed by all other Be X-ray binaries  
as initially established by Corbet (1986). 
The source has been extensively studied in outburst by Hulleman et  
al. (1998) with \B,  Baykal et al. (2000, 2002) with {\em RXTE},   
\.Inam et al. (2004) with {\em RXTE} and {\em XMM-Newton},  Reig et  
al. (2004) at optical and infrared wavelengths,  
and by Lutovinov et al. (2003), Sidoli et al. (2004) and Blay et  
al. (2004) with {\em INTEGRAL}.   
  
We present here for the first time spin phase resolved measurements  
at energies above 20 keV, based on hard X-ray  {\em INTEGRAL}   
observations of SAX J2103.5+4545 during a bright outburst.  
  
\section{Results}  
The data were obtained during the AO1 {\em INTEGRAL}  
observation  of the Cygnus region, performed from May 2 to May 3 2003  
(52761.32--52762.43 MJD).   
We use  data from the coded mask imager IBIS/ISGRI (Ubertini et  
al. 2003; Lebrun et al. 2003) with an exposure of 75.06  
ks, and from the JEM-X monitor (Lund et al. 2003) with an exposure   
time of 33.34 ks.     
The data reduction was performed using the standard software and 
response files provided in OSA 3.0 (\cite{gold03}).      
We extracted 32 channel spectra for IBIS/ISGRI and 256 channel spectra  
for JEM-X. The spectra were analyzed with the XSPEC V. 11.3 software  
package (\cite{Arnaud96}).    
The background subtracted ISGRI light curve for the timing analysis  
was reduced using  dedicated software ($ii\_light\_extract$,
  version to be included in OSA 5.0).   
The ISGRI light curve per energy band was obtained  through a  
simultaneous fit, for every 20 s long time bin, of all sources detected in  
the overall image and the background using the standard  
least squares method in the detector domain.  
For JEM-X the background subtracted light curve was obtained using   
standard software (see JEM-X Analysis User Manual).   
  
\subsection{X-ray Timing}  
  
We searched for coherent pulsations of the source in the 20--40 keV energy   
band where we have the best statistics.    
Using the 20 s binned light curve of the source in the 20--40 keV energy   
band, we computed a Power Density Spectrum (PDS) in the frequency 
range between   
5$\times$10$^{-5}$ and 0.025 Hz from Fast Fourier Transforms. In the  
resulting PDS an evident signal is present at  $\nu = 2.8296  
\times 10^{-3}$ Hz, and first and second harmonics are also  
visible. The highest peak in the power spectrum,   
the first harmonic, corresponds to a nominal period of 353.41 s.    
The accurate period was found using the phase-delay  
fitting method (see e.g.,\ \cite{nagase89}) resulting in  P = 353.45 
$\pm$ 0.16 s, with an upper limit to the pulse period derivative of 
$\mid$$\dot P$$\mid$ $<$   2.2 s yr$^{-1}$.   
In this section the errors are given at 1$\sigma$ confidence level for  
a single parameter. The measured pulse period is in agreement with the   
expected value using the pulse period of 354.794 s (52761.87 MJD) and 
the spin-up rate of $\sim$ 7.4 $\times$ $10^{-13}$ Hz s$^{-1}$ 
reported by \.Inam et al. (2004).

\begin{figure}[htb]  
   \centering  
   \includegraphics[width=6.5 cm, angle= -90]{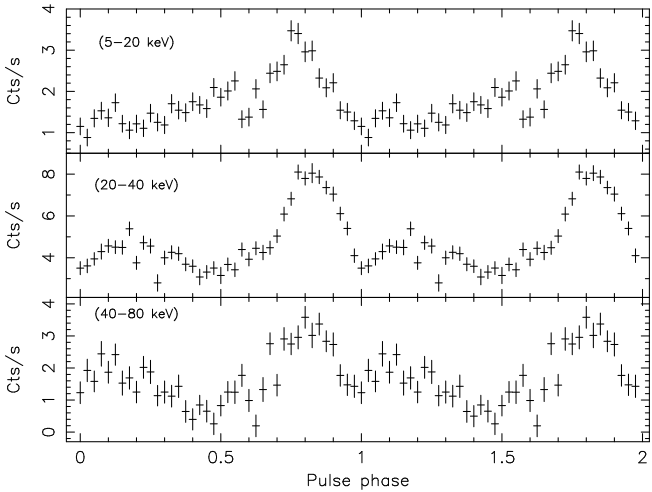}  
 \caption{JEM-X and IBIS/ISGRI light curves of SAX J2103.5+4545 folded at the  
   NS spin period. From top to bottom, in the energy range 5--20  
   keV (JEM-X), 20--40 keV and 40--80 keV (IBIS/ISGRI),  
   respectively. Each pulse profile is repeated once for clarity.}  
              \label{img:bestperiod}  
 \end{figure}  
Fig. \ref{img:bestperiod} shows the folded light curve at the best  
period. These profiles show a very prominent peak   
($\sim$ 51 s FWHM) around phase $\approx$ 0.8. The pulse fraction (PF),   
 defined here as PF = $(I_{max}-I_{min})/(I_{max}+I_{min})$, increases  
with energy, going from 47.5$\pm$0.3\% in the 5--20 keV energy   
band to  45.4$\pm$6.3\% at 20--40 keV and to 80.2$\pm$8.9\%   
at 40--80 keV. In the energy band 80--150 keV the low signal-to-noise   
ratio does not allow  a pulse fraction to be determined.   
The X-ray pulse profile of SAX J2103.5+4545 also shows an  
energy dependent secondary peak around phase $\approx$ 0.2, which  
becomes more evident at energies above 20 keV.  
Finally, note that the hick-ups in the top panel at phase 0.6 seem to   
be consistent with those observed in the pulse profile of this source   
by {\em RXTE/XMM-Newton} (e.g.,\ \.Inam et al. 2004). Note that these
hick-ups are also present in the 40--80 keV pulse profile (bottom
panel), while are not significantly present in the 20--40 keV pulse
profile (middle panel).   
  
\begin{table*}[htb]  
\begin{center}  
\caption{\label{spetab} Best fit parameters of the absorbed {\sc Comptt} model  
for the spectra of SAX J2103.5+4545 together with 90\% confidence errors   
for a single parameter.}  
\begin{tabular}{@{}lcccc}  
\hline  
Pulse phase             & 0.0 - 0.5 &  0.5 - 1.0 & All & All\\  
\hline   
$N_{\rm H}$ ($10^{22}\,{\rm cm}^{-2}$, fixed) & 3.45 & 3.45 & 3.45 & 3.45 \\   
$kT_{\rm 0}$ (keV) & 1.5 (f) & 1.5 (f) & 1.50 $\pm $ 0.17 & 1.5 (f)\\  
$kT_{\rm e}$ (keV) & 13.0 (f) & 13.0 (f) & 13.68 $\pm$ 1.85\ & 13.0 (f)\\  
$\tau$ & 2.08 $\pm$ 0.17 & 2.5 $\pm$ 0.13 & 2.37 $\pm$ 0.43 & 2.53
$\pm$ 0.14\\   
$\chi^2_{\rm red}$ & 1.2 (49 d.o.f.)& 1.2 (49 d.o.f.) & 1.09 (47
d.o.f.) & 1.06 (49 d.o.f.)\\  
$L_{5-200 keV}^{*}$ $(10^{36} {\rm erg}\, {\rm s}^{-1})$   & 2.95 &  4.64
& 4.17 & 4.21 \\  
\hline  
$^*$    Assuming a distance of 6.5 kpc.  
\vspace{-0.5 cm}  
\end{tabular}  
\end{center}  
\end{table*}  
  
\subsection{Spectral analysis and pulse phase resolved spectra}  
\label{sec:spec}  
 \begin{figure}[htb]  
 \centering  
   \includegraphics[width=5.0 cm, height=8 cm, angle=-90]{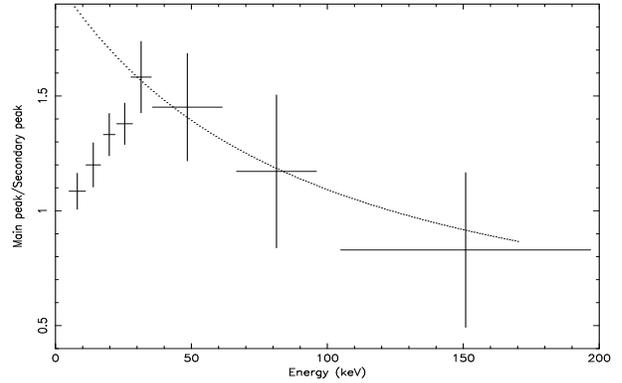}  
      \caption{Ratio of the ``main'' to ``secondary'' peak of the  
      pulse profile as a function of energy from 5 to 200  
      keV.  The dotted line is from the best fit two polar cap  
      diffusion model (see \S 3).  }  
         \label{fig_ratio}  
   \end{figure}  
We extracted the averaged over the whole observation spectrum of the   
source.   
In order to investigate the pulse phase dependence of the high energy  
emission from this source, we selected two phase intervals in the pulse  
profile of Fig. \ref{img:bestperiod}, one corresponding to the main peak   
(0.5--1.0) and the other corresponding to the secondary peak (0--0.5).   
From each of these phase intervals we extracted JEM-X and IBIS/ISGRI   
spectra.   
The spectral analysis has been carried out in the 5--20 keV energy range   
for JEM-X and in the 20--200 keV energy range for IBIS/ISGRI; the JEM-X   
spectra were rebinned to 28 channels to limit the oversampling of the 
instrumen resolution to a factor of 3.
  
The two phase-resolved spectra look different as is evident from their   
ratio plotted as a function of energy in Fig. \ref{fig_ratio}. The ratio   
gradually increases from 5 to 30 keV and then decreases at higher energies.  
Note that the Direct Continuum (DC) level is not subtracted from these  
spectra; this might introduce a variation of the spectral ratio simply  
due to the fact that the DC level changes with energy. In principle we could   
estimate the DC level from the spectrum at the minimum of the folded light   
curve, although we do not know if all the emission at the minimum of
the light curve is due to a DC level, or instead is due to unpulsed
emission from the polar caps (this is possible for instance if parts
of the polar caps are always visible). However, the statistics of
these data is not good enough to extract a meaningful spectrum at the
minimum of the folded light curve.   
Note that we took into account the presence of a DC level in the diffusion  
model that we used to fit this spectral ratio (see \S 3).  
  
  \begin{figure}[htb]  
 \centering  
   \includegraphics[width=9.0 cm, height=5 cm]{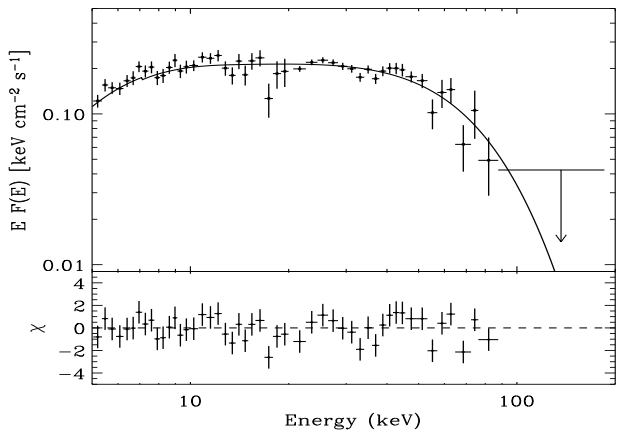}  
   \vspace{0.2 cm}  
      \caption{The 5--200 keV unfolded spectrum of SAX J2103.5+4545  
	along with the  best fit {\sc comptt} model.}  
         \label{fig_spec1}  
   \end{figure}  

To fit the total spectrum, we first used an
  absorbed cut-off power law model  ($\chi^2_{\rm red}=1.28$, 48
  d.o.f), were the photon index was 
found to be $\Gamma \sim 1.45$ with an cut-off at $E_{cut}
\sim 30$ keV. A better fit was found using an absorbed power law
multiplied by a high-energy exponential cutoff model ($\chi^2_{\rm
  red}$ = 1.37, 47 d.o.f.), where 
the parameters found were similar to the ones reported in Blay et
al. (2004).   
We then substituted this phenomenological model with a physically  
motivated thermal Comptonization model {\sc comptt}, using the slab
geometry (Titarchuk 1994), which gives a  $\chi^2_{\rm red}$ of 1.09
(47 d.o.f.).    
Usually the power law plus high-energy cutoff model is a good approximation  
of the spectra of high-magnetic field X-ray pulsars, due to the fact that  
the strong magnetic field modifies the Compton cross section giving a   
sharp cutoff in the spectrum. However, in our case, the {\sc comptt} model   
provides a slightly better description of the data with respect to the   
``empirical'' power law multiplied by a high-energy exponential cutoff
on the entire dataset at 81\% confidence level (estimated by means of
an F-test).   
We therefore prefer to report here the best-fit parameters of the   
{\sc comptt} model.   
We interpret the parameters of the {\sc comptt} model in the following way:  
at the bottom of the accretion column, i.e.\ at the hot spot on the NS
surface, a blackbody emission, with a temperature $kT_{\rm 0}$,
contributes the soft component of the spectrum, and provides also the
seed photons for the Comptonization (\cite{Inam04}). These are
Comptonized in the hot plasma present in the accretion column at a
temperature $kT_{\rm e}$ and optical depth $\tau$.  

In fitting the pulse phase resolved spectra,
we found that the spectral parameters of the {\sc comptt} model 
for the two spectra are compatible at 90 \% confidence level. 
In order to better constrain
the optical depth, $\tau$, for the simple difussion model (see Discussion)
we fixed the seed photons temperature $kT_{\rm 0}$ and the hot plasma
temperature $kT_{\rm e}$  to a single value (see Table 1).  
In Fig. \ref{fig_spec1}  we show the unfolded $E F(E)$ spectrum of the entire  
observation, plotted together with the residuals in units of $\sigma$ with  
respect to the best fit {\sc comptt} model. 
The best fit parameters together with the 90\% c.l.\ errors for a
single parameter are reported in Table 1.

\section{Discussion}  
\label{sec:concl}  
We analyzed for the first time the shape of the pulse profile of   
SAX J2103.5+4545 at energies above 20 keV.  
The observed X-ray pulse profile shows a broad main peak and a secondary   
peak that seems to become more prominent at higher energies. The 
differences between   
the spectra of the main (pulse phase: 0.5--1) and secondary (pulse   
phase: 0--0.5) peak are evident in their ratio, which increases with increasing  
energy up to $\sim$ 30 keV and then decreases, indicating that the main peak  
is characterized by an excess of emission in the energy range 20--50 keV;   
the difference in the mean 5--200 keV luminosity between the two pulse phase   
resolved  spectra is a factor of about 1.5.   
The pulsed fraction is observed to significantly and steeply increase  
with energy,  from $\sim$ 45\% up to 40 keV to $\sim$ 80\% in the  
range 40--80 keV. A similar behaviour was also found
 for the Be/X-ray transient RX J0117.6-733 where the pulsed percentage
 increased to at least 79\% in the 20-70 keV band (Macomb et al, 1999).
  
The average spectrum, as well as the pulse phase resolved spectra (i.e.\  
main peak spectrum and secondary peak spectrum) can be well described by   
a Comptonization model.  
According to this model, the seed photons for the Comptonization have  
a temperature of $kT_{0} \sim$ 1.5 keV, injected from the bottom of  
the accretion shock of electron temperature $kT_{\rm e}$ $\sim$ 13 keV  and  
Thomson optical depth $\tau  \sim$ 2.4. The electron temperature is  
too high for an accretion disk (\cite{frank}).     
  
In the hypothesis that the two peaks in the pulse profile come from   
the two polar caps on the NS surface, the variation of the pulse profile  
with energy cannot be simply explained by geometrical effects (the  
viewing angles of the polar caps, when directed towards     
the observer, are different and the projected emission areas in the line   
of sight are different) because the geometrical effects can change   
the observed flux but not the spectral shape. We therefore propose that  
the differences between the spectra at the two peaks is mainly caused  
by scattering effects.  Since the main peak do not vary significantly
with the energy we consider in  our model that only the second peak is
scattered.   

Considering that the radiation in the main peak comes from one polar cap  
and from the continuum of the NS, the flux can be written $I_{\rm 
  spot}+I_{\rm NS}$.    
The secondary peak then is $I_{\rm spot}'e^{-\tau'} + I_{\rm NS}$
(part of the secondary pole emission ($1 - e^{-\tau'}$) is scattered
away from the line of sight and does not contribute to the the
observed peak flux). $\tau'$ is given from the Comptonization fit
parameter and from the Klein-Nishina equation, noting that in our case
$kT_{\rm e} \ll mc^{2}$.    
Above 30 keV, the pulsed fraction is 80\% which shows that the continuum   
from the NS has a small contribution. In that energy range, the ratio given   
in Fig. \ref{fig_ratio} represents the ratio of the two flux peaks, which   
indeed is well fitted by the simple diffusive model, $(1+\alpha)/  
(1+\alpha'e^{-\tau'})$, where $\alpha=I_{\rm spot}/I_{\rm NS}$ and  
$\alpha'=I_{\rm spot}'/I_{\rm NS}$ (note that $I_{\rm spot}$ and
$I_{\rm spot}'$   
may be different due to geometrical effects, since the viewing angles of the  
magnetic poles, when directed towards the observer, may be different giving a   
difference in the projected emission areas along the line of sight).   
Although the statistics are not good enough to constrain all the parameters   
(the error bars on the single parameters are quite large), interestingly   
we always obtain that $\alpha' = 0.1 \alpha$. We can therefore quantify the  
geometrical difference between the two projected emission areas in 10\%.  
On the other hand, in the energy range between 5--30 keV with a pulsed   
fraction of $\sim$ 50\%, the continuum of the NS contributes half of the   
emission. The emission of the secondary peak is likely scattered through   
the Compton shock layer, above the hot spot, and contributes to the continuum.    
  
Finally, from the observed double-peaked folded light curve, we infer a  
limit on the angles that determine the geometry of the accretion powered   
pulsar (if the NS has a pure centered dipole field); these are the  
angle between the rotation axis and the direction to the observer, $i$, and  
the angle between the magnetic and rotation axes, $\beta$.  
Assuming a typical mass value of 20 M$_{\odot}$ for the BOV companion  
star (Reig et al. 2004) and a 1.4 M$_{\odot}$ NS, from the mass  
function the inclination angle of the system is $i \, \sim \,  
30^{\circ}$.  
The parameter space when two polar caps are seen is determined by the  
condition $i \, > 90^{\circ} - \beta - \delta g$, where $\delta g$ is the  
gravitational light deflection of the emitted X-ray  
(\cite{belo02}). In our case, considering $\delta g$ = 0, this leads  
to $\beta <$ $60^{\circ}$, or higher with non zero $\delta g$.  
The gravitational light bending effect is less then 10$^{\circ}$  
(see Bulik et al. 2003).

\acknowledgements  
MF is grateful to P. Laurent and G. Israel for valuable  
discussions and to A. Gros and S. Chazalmartin for providing the light  
curve software. This work is supported through the CNES and the CNRS 
"GrD Phenomenes Cosmiques de Haute Energie"

\end{document}